\documentstyle[prl,aps,twocolumn]{revtex}

\tightenlines   %for single spacing in preprint

\input{psfig}

\begin{document}
\draft

%\wideabs{

\twocolumn[
\hsize\textwidth\columnwidth\hsize\csname @twocolumnfalse\endcsname

\title{ \vspace{-2mm} \hfill {\small DFPD 99/TH/51, TRI-PP-99-34} 
\vspace{2mm}\\
Isoscalar off-shell effects in threshold pion production from $pd$ collisions}

\author{L.~Canton$^{1}$ and W.~Schadow$^{2}$}
\address{$^{1}$Istituto Nazionale di Fisica Nucleare, Sez. di Padova,
Via F. Marzolo 8, Padova I-35131 Italia
\ \\
$^{2}$TRIUMF, 4004 Wesbrook Mall
Vancouver, British Columbia Canada V6T 2A3}

\date{April 27, 2000}

\maketitle

\begin{abstract}
We test the presence of pion-nucleon isoscalar off-shell effects 
in the  $pd\rightarrow \pi^+ t$ reaction around the threshold region.
We find that these effects significantly modify the production cross section
and that they may provide the missing strength needed to reproduce the 
data at threshold. 
\end{abstract}

\vspace{6pt}
%}
]

 \section{Introduction}
Pion production  from nucleon-induced reactions has the potential
to probe the nuclear phenomena at short distance since it involves processes
transferring large momenta to the target nucleus. But the pion also {\em 
mediates} the nuclear force; hence meson production (or absorption) plays
a fundamental role in hadron dynamics because it may reveal facets of 
meson-baryon couplings,  and of meson-exchange processes in general, 
which would remain hidden otherwise.

In absence of reliable calculations on meson dynamics within 
an interacting multinucleon context,
one has to rely on the determination of the reaction mechanisms which 
dominate the process. Even so, the analysis of the process is complicated 
by the fact that a general treatment of the reaction mechanisms reveals
the occurrence of many terms, and one is forced to introduce
further assumptions in order to reduce the number of terms to a few, 
tractable ones. This reduction clearly introduces ambiguities making it 
more difficult to extract informations about the nuclear wave function
at short distances, or about the modifications of the hadron interactions
because of the presence of other nucleons. 
The situation is somewhat simplified if we consider nucleon-induced
production close to the pion threshold, since there the $s$-wave 
mechanisms of the elementary $NN$$\rightarrow\pi$$NN$ inelasticities 
dominate, while the $p$-wave mechanisms (including the isobar 
degrees of freedom) can be treated as corrections. 

In the past decade, with the aim to clarify the nature of the elementary
$NN$$\rightarrow \pi$$NN$ $s$-wave inelasticities, 
a great deal of experimental and theoretical
activity has been made in pion production from $NN$ collision at energies 
close to threshold. 
The situation has been recently reviewed by Meyer~\cite{Meyer98}.
Advances in experimental techniques allowed to measure
in particular the reaction $pp\rightarrow \pi^o pp$ cross section very close 
to threshold. This reaction filters the $s$-wave $\pi$$N$ coupling 
in the isoscalar  channel.
Standard theory including the one-body
term and isoscalar rescattering constrained by the $\pi$$N$ 
scattering lengths  underestimated the cross section by a factor of five.
Unexpectedly, there have been two theoretical explanations for this
discrepancy, not just one. The enhancement in the cross section can be  
explained by invoking short-range nucleon-nucleon effects \cite{Lee93},
where the important effects can be simulated by $\sigma$ and $\omega$ exchanges
\cite{Horowitz94}. The explanation is appealing, since in this case the 
pion field couples with the two-nucleon axial charge operators, and this 
coupling provides an explicit link to the inner part of the nucleon-nucleon 
interaction, which is notoriously difficult to disclose. But the results have 
been entirely explained  by resorting also to an off-shell enhancement
of the isoscalar $\pi$$N$ amplitude \cite{Hernandez95}.

The calculation by Hern\'andez and Oset employs 
an $s$-wave rescattering diagram where the isoscalar amplitude is 
described by the combined effect of a strong short-range repulsion 
and a medium range attraction of similar strenght, where the repulsion is 
represented by a contact term and the attractive part is parameterized 
by means of a $\sigma$-exchange diagram. Originally,
this representation has been derived 
with dispersion theoretic methods by analyzing the experimental data on 
pion-nucleon scattering with discrepancy functions 
\cite{Hamilton67}.
Subsequently, the fit in Ref.\cite{Hamilton67} has been
reinterpreted as being generated by a sigma-exchange term plus a 
$u$-channel term including $\bar N$ exchange 
(or virtual $N\bar N$ creation) and other short-range contributions
\cite{Maxwell80,Garcia-Recio88}. 
%Also Maxwell, Weise, and Brack 
%noticed the presence of an off-shell enhancement in the isoscalar 
%channel at threshold energies.

That $s$-wave pion production/absorption is governed by off-shell effects 
has been known for quite a few years\cite{Hachenberg78,Maxwell80}.
Hachenberg and Pirner used a field theoretical description for 
$\pi$$N$ scattering in $s$ wave based on the linear $\sigma$ model and on
pseudoscalar $\pi$$NN$ interaction. This combination results in a large 
cancellation between the $\sigma$-exchange diagram and the nucleon 
Born terms. The cancellation however breaks down producing an enhancement
when one pion leg is off-mass shell, as happens when the pion 
rescatters before being absorbed. 
On the contrary, in the isovector channel the cancellation occurs 
more efficiently off-shell, producing a suppression of the 
charge-exchange interaction. Thus, the relative importance of the 
isospin odd and even channels is reversed because of the off-shell 
extrapolation. 
Yet, another off-shell extrapolation has been considered in 
Ref.\cite{Hanhart95}, derived from a field theoretic model
of the $\pi$$N$ interaction which has been developed at J{\"u}lich
\cite{Schutz94}. This approach is similar in spirit to the model
developed by Hachenberg and Pirner, but here the meson exchanges in 
the scalar and vector channels are derived from correlated 2$\pi$ exchanges. 
The rescattering diagram of Ref.~\cite{Hanhart95} has a less 
pronounced off-shell enhancement with respect to both $\pi$$N$ models of 
Refs.\cite{Schutz94} and \cite{Hernandez95}, and one must introduce 
here two-nucleon short-range mechanisms such as those mediated by 
$\sigma$ and 
$\omega$ exchange in order to reproduce the total cross section for 
$pp\rightarrow pp\pi^o$ at threshold.
Finally, the problem of the off-shell extrapolation for the pion rescattering 
mechanism has been considered also in the more systematic framework
of chiral-perturbation theory ($\chi$PT) \cite{Cohen96}. 
The main feature here is that the rescattering diagram, once extrapolated 
off-shell, produce a negative interference with the one-body term, thus 
yielding a 
cross section substantially smaller than the measured ones. In this case the 
inclusion of heavy meson-exchange effect does not solve the discrepancy.
However, in another chiral-perturbation approach based on full 
momentum-space treatment\cite{Sato97}, the rescattering diagram was shown 
to be larger  by a factor of 3, thus leading to a substantial increase at 
threshold. More recently, the $\chi$PT approach has been carried further on 
by considering pion loop diagrams which might simulate $\sigma$-exchange 
phenomena, with the finding that these higher order contributions 
provide important improvements, but questioning at the same time
the convergence properties of the power counting expansion for the specific
reaction under consideration \cite{Gedalin99}.

The three-nucleon system is a richer testing ground for studies
of pion production and scattering. The addition of just one nucleon increases
the complexity of the reaction  which involves now the simplest
nontrivial  multinucleon system where it is possible to test the fundamental
$NN$$\rightarrow\pi$$NN$ process and, at the same time, to solve the 
accurately the nuclear dynamics with Faddeev methods.
Applications of Faddeev methods to pion production/absorption started very 
recently with studies centered around the $\Delta$ resonance 
\cite{Kamada97,Canton97} and herein we apply the same technique of
Ref.~\cite{Canton97} to study pion production at threshold.
Besides the obvious difficulty of performing calculations with three 
nucleons instead of two, one encounters here the additional difficulty
that for the $pd\rightarrow\pi^+ t$ reaction one must include from the start
the effect of $p$-wave mechanisms, on top of the $s$-wave ones. This contrasts 
with the findings for the two-nucleon case, where the effect of the $p$-wave
mechanisms (including the $\Delta$), tends gradually to zero in approaching 
the  threshold limit. The effect of this difference can be immediately 
perceived in the differential production cross section, 
since for $NN$ collision the angular
dependence evolves gradually with energy, while in the case of $pd$
collisions it exhibits a remarkable $s$- and $p$-wave interference in the 
threshold region, with strong forward-backward asymmetry\cite{Nikulin96}.

In this work, we have centered our study on the effects due to the
$s$-wave rescattering processes, taking into account both isoscalar
and isovector components and their interference to the
$p$-wave mechanism (containing also the $\Delta$ degrees of freedom).
We have in particular taken into account the off-shell effects
in the isoscalar channel by following the same prescription 
suggested in Ref.~\cite{Hernandez95} to explain the size of the excitation
function for the $pp\rightarrow pp\pi^o$ process. It is important to stress 
that there are still large uncertainties inherent to this 
off-shell extrapolation, and such calculations should be repeated 
possibly also with other off-shell extensions. Moreover, other
production mechanisms here omitted should be possibly included in 
the calculation, at least those mechanisms which proved to have relevant 
interference effects in $NN$ collisions. But to implement the 
production mechanisms in a three-nucleon process is not a trivial
task and needs to be done gradually.
 
At present stage, where we believe that the importance of the 
off-shell effects in $s$-wave pion production from $NN$ collisions
has been fairly well established by various groups,  
it is clearly of great relevance to consider the consequences of such 
effects on more complex reactions. Herein we provide the results 
obtained when calculating off-shell effects in $pd$ collisions.

\section{theory}

The production mechanisms are constructed starting from
the following effective pion-nucleon couplings:
\begin{eqnarray}
\label{Lagrangiana}
{\cal L}_{\rm int}
& = & 
{f_{\pi NN}\over m_\pi} 
\bar\Psi\gamma^\mu\gamma^5\vec\tau\Psi \cdot \partial_\mu \vec \Phi \\
&&
-4\pi{\lambda^{}_I\over m_\pi^2}
\bar\Psi\gamma^\mu\vec\tau\Psi \cdot 
\left[\vec\Phi \times \partial_\mu \vec \Phi\right] 
-4\pi{\lambda^{}_O\over m_\pi}
\bar\Psi \Psi 
\left[\vec\Phi \cdot \vec \Phi\right] . \nonumber
\end{eqnarray}

\noindent
The first term represents the gradient coupling to the isotopic
axial current, while the second denotes the coupling to the isovector
nucleonic current, and the last is the pion-nucleon coupling in the 
isoscalar channel. 

As is well known\cite{Ericson88}, a good guess for the coupling 
constants can be obtained from chiral symmetry and PCAC,
which constrain  the three constants to be of the order
\begin{equation}
f_{\pi NN} /m_\pi \simeq g_A/(2f_\pi),
\end{equation}
\begin{equation}
4\pi \lambda^{}_I/m_\pi^2 \simeq 1/(4f_\pi^2),
\end{equation}
\begin{equation}
4\pi \lambda^{}_O/m_\pi \simeq 0 ,
\end{equation}
where $g_A$ ($\simeq 1.255$) is the axial nucleonic normalization,
and $f_\pi$ is the pion decay constant ($\simeq 93.2$ MeV).
The first condition follows directly from the Goldberger-Trieman
relation, while the last two are implied by the Weinberg-Tomozawa ones.
Current phenomenological values for $f_{\pi NN}^2/(4\pi)$ can reach 
values as low as 0.0735 \cite{Arndt90}, 0.0749 \cite{Stoks94},
or 0.076 \cite{Arndt95}, 
but also values around  0.081 \cite{Ericson95}
have been considered acceptable. Some years ago common values
were centered around 0.078--0.079 \cite{Koch80,Bugg73}.
Similarly, from the pion-nucleon scattering lengths, $\lambda^{}_I$ is 
determined within the range 0.055--0.045, while the weaker 
isoscalar coupling has typically larger indetermination, 
ranging from 0.007 to $-0.0013$~\cite{Lee93,Hernandez95}. The
isovector and isoscalar couplings,
when combined with the axial $\pi$$NN$ vertex, are the basic ingredients 
for the two-nucleon $s$-wave rescattering mechanisms, while the 
axial-current coupling alone 
forms the basis for the one-body production process.

The matrix elements for the rescattering process require an 
off-mass-shell extrapolation of the two constants $\lambda^{}_I$ and 
$\lambda^{}_O$, since the rescattered pion can travel off-mass-shell.
For $\lambda^{}_O$ we consider the off-shell structure 
previously employed in the $pp\rightarrow pp\pi^o$ process by 
Hern\'andez and Oset (Ref.~\cite{Hernandez95}) which is 
based on a parametrization
due to Hamilton~\cite{Hamilton67},
namely
\begin{eqnarray}
\lambda^{}_O(\tilde{q},\tilde{p})&=&\lambda_O^{\rm on} g^{}_O(\tilde{q},
\tilde{p})
\\
&=&-{1\over 2}(1+\epsilon) \, m_\pi\left(a_{sr}+a_\sigma 
{m_\sigma^2\over m_\sigma^2-(\tilde{q}-\tilde{p})^2}\right),  \nonumber
\end{eqnarray}
with $m_\sigma=550$ MeV, $a_\sigma=0.22\ m_\pi^{-1}$, 
$a_{sr}=-0.233\ m_\pi^{-1}$, and $\epsilon=m_\pi/M$.
In the threshold limit,
$(\tilde{q}-\tilde{p})^2\simeq (\tilde{q}_o-m_\pi)^2-
{\bf \tilde{q}}^2$, where $\tilde q$ is the transferred
4-momentum between the two active nucleons. According to
previous treatments of the 2$N$ $\pi$-production amplitude, 
the time-component $\tilde  q_o$ 
is fixed around $\tilde q_o\simeq  m_\pi/2$, while the space
component $\tilde  {\bf q}$ represents a loop variable and is integrated 
over.

This form leads to an on-shell value of the order of
0.007. The on-shell value derives from a cancellation between
the short-range term, $a_{sr}$, and the $\sigma$-exchange term, $a_\sigma$,
while off shell the cancellation occurs only partially and thus leads
to the off-shell enhancement.
The use of a fictious sigma-exchange model should not be considered a 
crucial aspect of the model, since similar forms (possibly summed over a 
``distribution" of masses $m_\sigma$) can be easily obtained also 
in theoretical approaches based on subtracted dispersion relations.
The approaches based on subtracted dispersion relations such as 
$\pi$$N$ model of Ref.~\cite{Schutz94} lead indeed to 
similar off-shell enhancement.

For $\lambda^{}_I$ the extrapolation can be accomplished by invoking
$\rho$-meson dominance and the related 
Riazuddin-Fayyazuddin-Kawarabayashi-Suzuki identity,
which implies (on shell, for $\omega_\pi\rightarrow m_\pi$)
\begin{equation}
\label{eqlambda}
{\lambda_I^{\rm on}\over m_\pi^2}={f_\rho^2\over 8\pi m_\rho^2},
\end{equation}
with the corresponding off-shell extrapolation
\begin{eqnarray}
\lambda^{}_I(\tilde{q},\tilde{p})&=&\lambda_I^{\rm on} g^{}_I(\tilde{q},
\tilde{p}) \nonumber  \\ & = &
\lambda_I^{\rm on}{m_\rho^2\over m_\rho^2-(\tilde{q}-\tilde{p})^2}
\left({\Lambda_\rho^2\over\Lambda_\rho^2-(\tilde{q}-\tilde{p})^2}
\right)^2,
\end{eqnarray}
\noindent
where again we use $(\tilde{q}-\tilde{p})^2=(\tilde{q}_o-m_\pi)^2-
{\bf \tilde{q}}^2$ in the threshold limit.

The production matrix-elements in the nonrelativistic 3$N$ space
with such couplings are the following:
\begin{eqnarray}
\label{a1}
\langle 3N| \lefteqn
{ A^{\rm 1B} |3N,\pi\rangle =  % &  \nonumber\\ & 
{-if_{\pi NN}\over m_\pi} \,
{\mbox{\boldmath $\sigma$}_2 {\bf \tilde{p}}}
%\over (2\pi)^{3\over 2}\sqrt{2\omega_\pi}} 
[\mbox{\boldmath $\tau$}_2]_1^{z_\pi}}  \nonumber \\
& & \times\, \delta({\bf p'}-{\bf p}-{6+2\epsilon\over 
6(2+\epsilon)}{\bf P}_\pi)
\, \delta({\bf q'}-{\bf q}+{1\over 3}{\bf P}_\pi) 
\end{eqnarray}
for the one-body process,
\begin{eqnarray}
\label{a2o}
\langle 3N| \lefteqn
{ A_O^{\rm 2B} |3N,\pi\rangle = %& \nonumber \\ & 
\, 2i \, {f_{\pi NN}4\pi\lambda^{}_O \over m_\pi^2} }  \nonumber \\
& & \times \,
%\over (2\pi)^{3\over 2}\sqrt{2\omega_\pi}} 
\mbox{\boldmath $\sigma$}_3 
{\bf\tilde{q}}[\mbox{\boldmath $\tau$}_3]_1^{z_\pi}
{v_{\pi NN}(\tilde q)g^{}_O(\tilde{q},\tilde{p})\over m_\pi^2+\tilde{\bf q}^2 
-\tilde{q_0}^2 } \,
\delta({\bf q'}-{\bf q}+{1\over 3}{\bf P}_\pi)
\end{eqnarray}
\noindent
and
\begin{eqnarray}
\label{a2i}
\langle 3N| \lefteqn{
 A_I^{\rm 2B} |3N,\pi\rangle = %& \nonumber\\& 
{\sqrt{2}i \, {f_{\pi NN}4\pi\lambda^{}_I \over m_\pi^3}}} \nonumber \\
%\over (2\pi)^{3\over 2}\sqrt{2\omega_\pi}}
 & & \times \mbox{\boldmath $\sigma$}_3 {\bf\tilde{q}}
[\mbox{\boldmath $\tau$}_3
\times\mbox{\boldmath $\tau$}_2]_1^{z_\pi}
{v_{\pi NN}(\tilde q) g^{}_I(\tilde{q},\tilde{p})
\over m_\pi^2+\tilde{\bf q}^2-\tilde{q_0}^2 } \,
(\tilde{q}_0+\omega_\pi) \nonumber \nonumber \\
& & \times \,\delta({\bf q'}-{\bf q}+{1\over 3}{\bf P}_\pi)
\end{eqnarray}
for the two-body isoscalar and isovector rescattering, respectively.
$v_{\pi NN}(\tilde{q})$ is the hadronic form factor
of the $\pi$$NN$ vertex, whose structure is governed by the 
monopole-type cutoff $\Lambda_\pi$.
The momenta ${\bf p}$ and ${\bf q}$ are Jacobi momenta
for nucleon 2 in the (2+3) center-of-mass (c.m.), and nucleon 1 in the 
(1+2+3) c.m., respectively, while ${\bf P_\pi}$ is the pion momentum
in the total c.m.  Similarly, ${\bf p'}$ and ${\bf q'}$
are the Jacobi momenta for the three nucleons 
in the no-pion case. Other relevant pion momenta
are
\begin{equation}
{\bf \tilde{p}} \simeq {(3+\epsilon)\over 3(1+\epsilon)} {\bf P_\pi}
\end{equation}
and
\begin{equation}
{\bf \tilde{q}} \simeq  {\bf p} + 
{(6+2\epsilon)\over 6(2+\epsilon)} {\bf P_\pi} -{\bf p'} \, .
\end{equation}
In the actual calculation ranging from threshold up to the 
$\Delta$ resonance the on-shell couplings 
are further corrected by means of an Heitler-type (or $K$-matrix) form
($\eta$ is the pion momentum in pion-mass units

\begin{equation}
\hat\lambda^{}_O\simeq
{2\over 3}{\lambda^{}_O+\lambda^{}_I\over 1+2i\eta (\lambda^{}_O+\lambda^{}_I)}+
{1\over 3}{\lambda^{}_O-2\lambda^{}_I\over 1+2i\eta (\lambda^{}_O-2\lambda^{}_I)}
\, ,\end{equation}
\begin{equation}
\hat\lambda^{}_I\simeq
{1\over 3}{\lambda^{}_O+\lambda^{}_I\over 1+2i\eta (\lambda^{}_O+\lambda^{}_I)}-
{1\over 3}{\lambda^{}_O-2\lambda^{}_I\over 1+2i\eta (\lambda^{}_O-2\lambda^{}_I)}
\, .\end{equation}

This reduces the rescattering contributions at higher energies
but the correction is uninfluential in the threshold limit. 
On top of these processes, we have included also the two-body
mechanism mediated by $\Delta$ rescattering,
\begin{eqnarray}
\label{a2d}
\langle 3N| \lefteqn
{A^{\rm 2B}_\Delta |3N,\pi\rangle = %& \nonumber\\ & 
{-if_{\pi N\Delta}\over m_\pi} }  \nonumber \\
%{1\over (2\pi)^{3\over 2}\sqrt{2\omega_\pi}} 
& & \times \, {V_{N\Delta}({\bf p}',{\bf p}_\Delta) 
\mbox{\boldmath $ S$}_2  {\bf\tilde{p}} [\mbox{\boldmath $T$}_2]_1^{z_\pi}
\over T_{\rm cm}+M-{\cal M}_\Delta+p_\Delta^2/2\mu_\Delta
+{q'}^2/2\nu_\Delta
} \, \nonumber \\
& & \times \, \delta({\bf q'}-{\bf q}+{1\over 3}{\bf P}_\pi),
\end{eqnarray}
since its contribution becomes soon important as the energy increases.
The intermediate $\Delta$ momentum is defined as
\begin{equation}
{\bf p}_\Delta\simeq {\bf p}+
{(6+2\epsilon)\over 6(2+\epsilon)} {\bf P_\pi}\, .
\end{equation}
In Eq. (\ref{a2d}) $\mu_\Delta$ 
is the reduced mass of the $\Delta$-$N$ system,
while $\nu_\Delta$ is the reduced mass for the
$N$-($\Delta$$N$) partition. $T_{\text {c.m.}}$ is the c.m.
kinetic energy of the three nucleons in the initial state.
The $\Delta$$N$ transition potential is determined by 
the $\pi$ plus  $\rho$ exchange diagrams, where
the pseudoscalar meson 
provides the typical longitudinal structure to the $\Delta$$N$
transition, i.e.,
$(\mbox{\boldmath$\sigma$}_3\cdot  \bf{\tilde{q}})
(\mbox{\boldmath$S$}^\dagger_2\cdot  \bf{\tilde{q}}) 
(\mbox{\boldmath$\tau$}_3\cdot\mbox{\boldmath$T$}^\dagger_2)$,
while the vector meson generates the transversal contribution
$(\mbox{\boldmath$\sigma$}_3\times  \bf{\tilde{q}})
(\mbox{\boldmath$S$}^\dagger_2\times  \bf{\tilde{q}}) 
(\mbox{\boldmath$\tau$}_3\cdot\mbox{\boldmath$T$}^\dagger_2)$.
${\bf S}_2$ (${\bf T}_2$) is the spin (isospin) operator
for the $\Delta\leftrightarrow$ $N$ transition.
For complete details on the employed transition potential,
and for other aspects connected with the isobar mechanism,
such as,  e.g., the detailed expression for the complex 
$\Delta$ mass ${\cal M}_\Delta$, we refer to a set of 
studies performed around the resonance region 
\cite{Canton98,Dortmans97,Canton96}. 
All amplitudes, Eqs. (\ref{a1}),(\ref{a2o}),(\ref{a2i}), and (\ref{a2d})
must be multiplied in addition by
the common factor $1/\sqrt{(2\pi)^3 \,2\omega_\pi}$. Moreover, taking into
account Pauli identity, the full
one-body mechanism results by multiplying Eq. (\ref{a1}) by the multiplicity
factor $\sqrt{3}(1+P)$, while the remaining two-body mechanisms
are multiplied by $2\sqrt{3}(1+P)$, where $P$ is the ternary  permutator 
which commutes the 3$N$ coordinates  cyclically/anticyclically.
Combining the $P$ operator with the given mechanisms in 3$N$ partial
waves is not a trivial task, and numerical treatment of the resulting
amplitudes has been a challenge.

The matrix elements are calculated between in and out nuclear states,
where the out-state is specified by the 
three-nucleon bound-state wave function (plus a free pion wave),
and the incoming state is determined by the deuteron-nucleon
asymptotic channel. For the 3$N$ bound state we have taken the triton 
wave function as has been developed, tested and calculated in 
Ref.\cite{Schadow98}.
As two-body input for the three-nucleon equations we used a separable
representation \cite{Haidenbauer} of the Paris interaction.  This form
represents an analytic version of the PEST interaction, originally
constructed and applied by the Graz group\cite{Haidenbauer84}.

We have in addition calculated the relevance of the three-nucleon
dynamics in the initial state (ISI) by solving the Faddeev type
Alt-Grassberger-Sandhas (AGS) equations \cite{Alt67}.  The AGS
equations for neutron-deuteron scattering go over into effective
two-body Lippmann-Schwinger equations \cite{Alt67} when representing
the input two-body {\em T}-operators in separable form. The {\em T} is
represented in separable form using again the above mentioned EST
method.  Applying the same technique to the $\pi$ absorption process,
an integral equation of rather similar structure is obtained for the
corresponding amplitude. The only difference is that the driving term
of the {\it n-d} scattering equation ( i.e., the
particle-exchange diagram, the so--called ``Z'' diagram) is replaced
by the off-shell extension of the plane-wave pion-absorption
amplitude.  More details can be found in \cite{Canton97,Schadow99} and
references therein.

\section{Results}

\begin{figure}
\centerline{\psfig{file=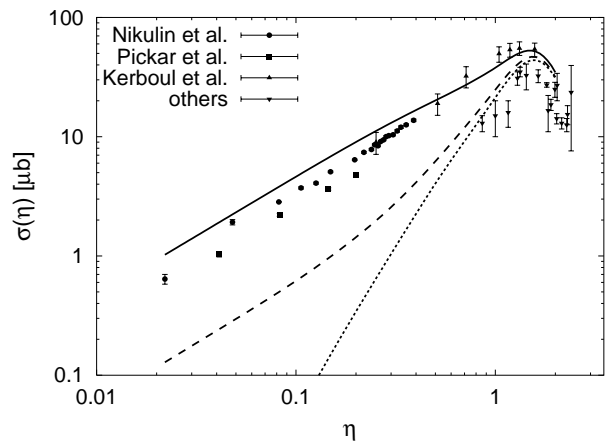,width=8.5cm,angle=-90}}
\vspace{3mm}
\caption{\label{figone} 
Production cross section for the $pd\rightarrow\pi^+t$ versus $\eta$.
The dotted line contains the sole $p$-wave mechanisms.
The dashed line includes also $\pi$$N$ $s$-wave rescattering mediated
by $\rho$-exchange. The solid line considers in addition the 
isoscalar off-shell effects. The data are from 
Refs.~[9,25,26].
%\cite{Nikulin96,Pickar92,Kerboul86}. 
The remaining data 
(``others") have been extracted from a world collection as explained in 
Ref.~[8].
%\cite{Canton97}.
}
\end{figure}

To exhibit the relevance of isoscalar off-shell effects for 
$pd\rightarrow\pi^+t$ we have calculated the integral 
cross section near threshold up to 
the $\Delta$ region. The calculated
results are compared with a collection
of data from Refs.~\cite{Nikulin96,Pickar92,Kerboul86} and 
others as explained in Fig.~\ref{figone}. Practically all the 
experimental data near threshold 
refer in fact to $\pi^o$ production from $pd$ collisions, and the 
comparison has been made assuming isospin invariance and hence 
implying a factor of 2 between the two cross sections. In so doing we have 
avoided the need to include the effects of Coulomb distortions 
in the exit channel.
Given the complexity of the
reaction dynamics which depends upon 
\linebreak

\noindent

\begin{figure}
\centerline{\psfig{file=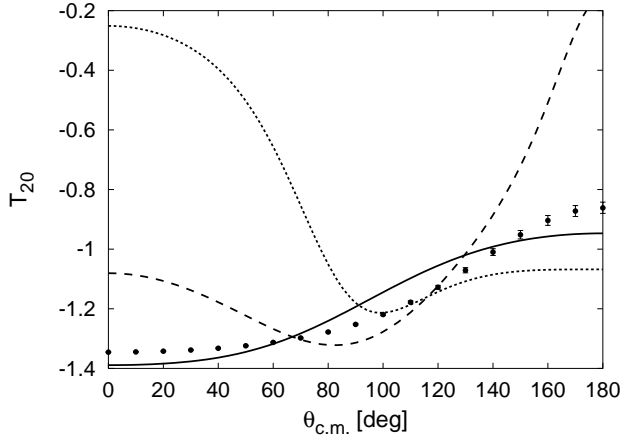,width=8.5cm,angle=-90}}
\vspace{2mm}
\caption{\label{figtwo} The deuteron tensor analyzing power
$T_{20}$ for $\eta=0.25$. The lines show the same calculations
as in Fig. 1. The points are extracted from Ref.~[9].
%\cite{Nikulin96}.
}
\end{figure}

\vspace{-1mm}

\begin{figure}
\centerline{\psfig{file=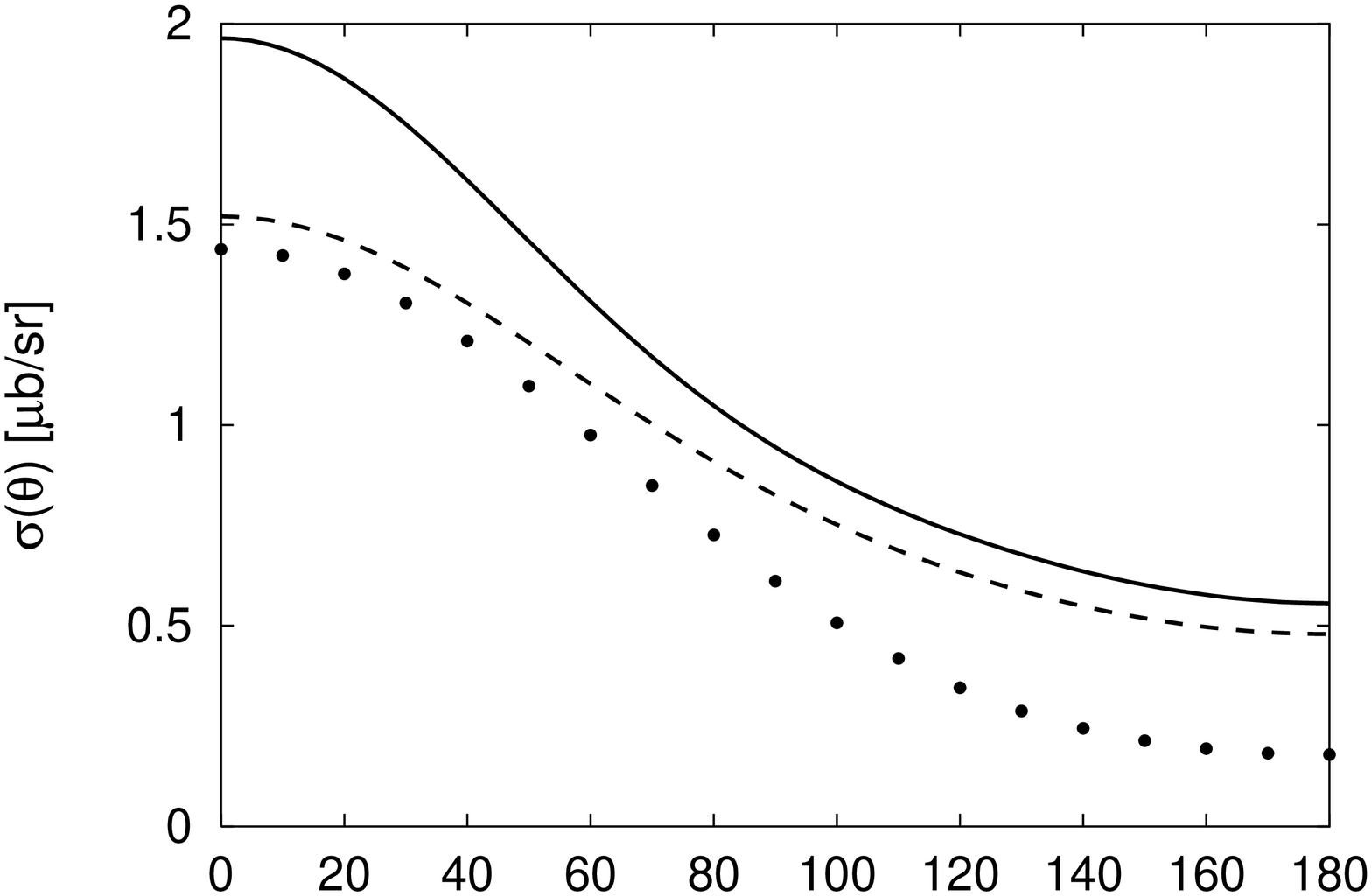,width=8.5cm,angle=-90}}
\vspace{-2mm}
\centerline{\psfig{file=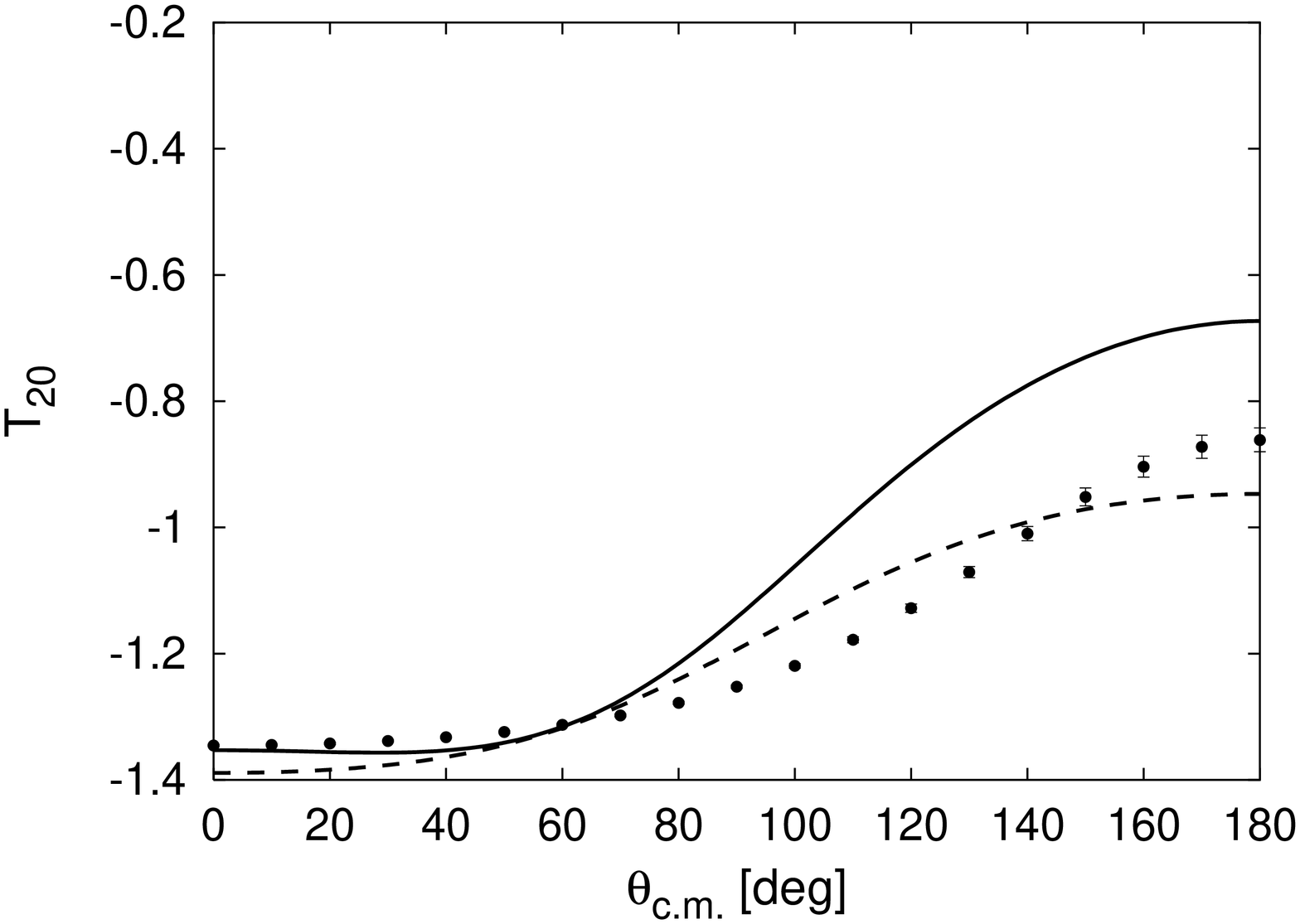,width=8.5cm,angle=-90}}
\vspace{2mm}
\caption {\label{figtre} 
Effect of 3$N$ initial-state correlation at $\eta=0.25$. Differential
cross section ($T_{20}$) on the top (bottom) panel.  In both cases
the solid line includes the 3$N$ ISI effects, while the dashed line
has been obtained in plane-wave approximation. 
The points are extracted from Ref.~[9].
%\cite{Nikulin96}.
}
\end{figure}

\noindent
several contributions, the isoscalar 
effects have been calculated on top of the other mechanisms we had
considered.
As explained previously, the model includes also 
$p$-wave $\Delta$ rescattering,
the one-body $p$-wave term, and the isovector $\rho$-exchange mechanisms.
The relevant parameters employed herein (cutoffs, coupling constants) 
have 
been tested previously against the $pp\rightarrow \pi^+ d$ 
process in Ref.~\cite{Canton98}. % Canton, Davini, Dortmans. 
For the $\rho$-exchange model we use $\lambda^{}_I=0.045$
in Eq.~(\ref{eqlambda}).
The results indicate that the modifications
 introduced by the isoscalar
contributions are significant over the whole range 
considered, and that the effect is one of the most pronounced at threshold.

Further evidences come from the results exhibited in Fig. \ref{figtwo},
where the deuteron tensor analyzing power $T_{20}$ is shown. 
Details on the formalism for the calculation of polarization observables can
be found in Ref. \cite{Canton98b}.
The points represent a best-fit to experimental data as given in 
Ref.\cite{Nikulin96}.  
The trend of the data is 
reproduced once both the isovector {\it and} isoscalar terms are
taken into account.

It is clearly important to assess the role of the initial 
state correlations, since the 
three-body dynamics between the nucleon and the deuteron  
could modify the whole picture
and undermine the conclusions of this work. For this reason,
we have calculated the ISI effects by solving the AGS
equations for the 3$N$ system using as input a separable representation
of the Paris interaction. The same Faddeev-like technique herein employed 
has been applied previously to pion production at the $\Delta$ resonance 
in Ref.~\cite{Canton97}.
In Fig. \ref{figtre}
one can examine the role of the 3$N$ dynamics in the initial state
from the angular dependence of the unpolarized production cross section 
and from $T_{20}$, for $\eta=0.25$. The modifications introduced by the 3$N$ 
dynamics are sizable, but the overall picture does not change drastically.
In addition, the 3$N$ effects improve the angular dependence
of both observables, thus possibly confirming 
our findings about the importance of the isoscalar off-shell effects. 

\acknowledgements 
The work of L.C. was
supported by the ``Ministero dell'Universit\'a e della
Ricerca Scientifica e Tecnologica" under the PRIN Project 
``Fisica Teorica del Nucleo e dei Sistemi a Pi\'u Corpi". 
The work of W. Sch. was supported by the Natural Science and 
Engineering Research Council of Canada.

\end{document}